\documentclass[twocolumn]{jpsj2}

\title{Antiferromagnetic Heisenberg chains with bond alternation and quenched 
disorder}

\author{Yu-Cheng \textsc{Lin}$^{1}$, Heiko \textsc{Rieger}$^{2}$ and 
        Ferenc \textsc{Igl\'oi}$^{3,4}$}

\inst{$^{1}$
  Institut f\"ur Physik, WA 331, Johannes Gutenberg-Universit\"at,
  55099 Mainz, Germany \\
$^{2}$
  Theoretische Physik, Universit\"at des Saarlandes,
  66041 Saarbr\"ucken, Germany \\
$^{3}$
 Research Institute for Solid State Physics and Optics,
 H-1525 Budapest, P.O.Box 49, Hungary \\
$^{4}$
 Institute of Theoretical Physics,
 Szeged University, H-6720 Szeged, Hungary}

\abst{We consider $S=1/2$ antiferromagnetic Heisenberg chains with alternating
 bonds and quenched disorder, which represents a theoretical model of
 the compound ${\rm Cu Cl}_{2x}{\rm Br}_{2(1-x)}(\gamma-{\rm
 pic})_2$. Using a numerical implementation of the strong disorder
 renormalization group method we study the low-energy properties of
 the system as a function of the concentration, $x$, and the type of
 correlations in the disorder. For perfect correlation of disorder the
 system is in the random dimer (Griffiths) phase having a
 concentration dependent dynamical exponent. For weak or vanishing
 disorder correlations the system is in the random singlet phase, in
 which the dynamical exponent is formally infinity.  We discuss
 consequences of our results for the experimentally measured
 low-temperature susceptibility of ${\rm Cu Cl}_{2x}{\rm
 Br}_{2(1-x)}(\gamma-{\rm pic})_2$.}

\kword{${\rm Cu Cl}_{2x}{\rm Br}_{2(1-x)}(\gamma-{\rm pic})_2$, low-temperature susceptibility, random Heisenberg chain, correlated disorder, strong disorder
renormalization group, random singlet phase, quantum Griffiths phase}

\begin{document}
\maketitle

\section{Introduction} 
Low-dimensional quantum systems (antiferromagnetic spin chains, ladders, 
two-dimensional systems, etc.) are fascinating objects which have been 
investigated intensively in experimental and theoretical works. From the
theoretical point of view these systems exhibit several unusual properties, 
for example quasi-long-range order, topological string order, a spin liquid 
state, quantum phase transitions, etc.. Many of these features can be 
illustrated by the $S=1/2$ antiferromagnetic Heisenberg chain, 
which is defined by the Hamiltonian:
\begin{equation}
H=\sum_i 2 J_i {\mib S}_i \cdot {\mib S}_{i+1}\;,
\label{hamilton}
\end{equation}
in terms of the spin-$1/2$ operators, ${\mib S}_i$, at site $i$. The
homogeneous model with $J_i=J$ is gapless and there is
quasi-long-range order in the ground state, i.e. spin-spin
correlations decay algebraically,\cite{lutherpeschel} $\langle
S^{\tau}_i S^{\tau}_{i+r}\rangle \sim r^{-1}$.  Here $\tau=x,y,z$ and
$\langle \dots \rangle$ stands for the ground-state expectation
value. Introducing bond alternation such that $J_i=J$ at $i={\rm
even}$ and $J_i=\alpha_d J$, ($\alpha_d >0$) at $i={\rm odd}$, the
system with $\alpha_d \ne 1$ is in the dimerized
phase,\cite{crossfisher} in which spin-spin correlations are
short-ranged and the excitation spectrum has a finite gap.  In the
bond alternating model the quantity $\delta_d=\ln \alpha_d$ serves as
a quantum control parameter and the quantum critical point is located
at $\delta_d^c=0$.

Quenched (i.e. time independent) disorder has a profound effect on the
low-temperature/low-energy properties of quantum
systems,\cite{sg-review} both at the quantum critical point and in an
extended region of the off-critical regime--the so called quantum
Griffiths phase\cite{griffiths,mccoy}. In the disordered version of
the uniform $S=1/2$ antiferromagnetic Heisenberg chain defined in
eq.(\ref{hamilton}), the couplings, $J_i$, are independent and
identically distributed random variables. A detailed study by
Fisher\cite{fisher} using an asymptotically exact strong disorder
renormalization group (SDRG) method\cite{MDH} revealed that the ground
state of the random model is the so called random singlet phase, in
which singlets are formed between spin pairs which could be
arbitrarily far apart. Average spin-spin correlations, which are
dominated by rare regions, are quasi-long-ranged: $[\langle S_i
S_{i+r}\rangle]_{\rm av}
\sim r^{-2}$, where $[ \dots ]_{\rm av}$ stands for averaging over
quenched disorder.  On the other hand, {\it typical} correlations are
much weaker and decay asymptotically as $\ln(\langle S_i
S_{i+r}\rangle_{\rm typ}) \sim r^{1/2}$. In addition, the dynamical
scaling in the random singlet phase is anomalous. The length scale 
$L$ and the energy scale $\Delta$, measured by the lowest gap,
are related logarithmically:
\begin{equation}
\ln \Delta \sim L^{\psi}, \quad \psi=1/2\;.
\label{gap_rs}
\end{equation}
In the random bond alternating Heisenberg model bonds at even ($J_{\rm e}$) 
and odd ($J_{\rm o}$) sites are taken from different distributions and the 
quantum control parameter is defined as
\begin{equation}
\delta=[\ln J_{\rm e}]_{\rm av}-[\ln J_{\rm o}]_{\rm av}\;.
\label{delta}
\end{equation}
For $\delta \ne 0$ the model is in the random dimer phase, in which
spatial spin-spin correlations are short ranged.\cite{HYBG96} Within a
range of finite dimerization $|\delta| < \delta_G>0$, dynamical
correlations are however still quasi-long-ranged, which is related to
the fact that the system is gapless. This region is called Griffiths
phase. In a finite chain of length $L$ the typical gap scales
asymptotically as:
\begin{equation}
\Delta \sim L^{-z}\;.
\label{gap_rd}
\end{equation}
Here the dynamical exponent, $z$, is a continuously varying function of the 
control parameter, $\delta$. On the border of the Griffiths phase, 
$\delta=\delta_G$, the dynamical exponent vanishes, whereas close to the 
random singlet phase it diverges with $\delta\to0$ as\cite{fisher}:
\begin{equation}
z \sim 1/\delta\;.
\label{z_delta}
\end{equation}
In the random singlet phase as well as in the Griffiths phase, thermodynamic 
quantities such as the low-temperature susceptibility, $\chi(T)$, the 
low-temperature specific heat, $C(T)$, and the low-field magnetization,
$M(h)$, at zero temperature are singular. For example the susceptibility 
in the random singlet phase behaves as:
\begin{equation}
\chi(T) \sim \frac{1}{T (\ln T)^2}\;,
\label{chi_rs}
\end{equation}
whereas in the Griffiths phase it is given by:
\begin{equation}
\chi(T) \sim T^{-1+\beta},\quad \beta=1/z\;.
\label{chi_rd}
\end{equation}
%
%Concerning another (more complicated) random Heisenberg spin systems we note, that a %random singlet phase has been found only at special points of one-dimensional and %quasi-one-dimensional (i.e. ladder) systems and often only for strong enough disorder, %c.f. for higher spin antiferromagnetic Heisenberg chains. Otherwise non-frustrated %systems are usually in a Griffiths-type phase, whereas in frustrated systems usually %large-spin formation happens during renormalization.

Recent experimental studies on the compound ${\rm Cu Cl}_{2x}{\rm
Br}_{2(1-x)}(\gamma-{\rm pic})_2$ show another type of competition
between bond alternation and randomness.\cite{ajiro,wakisaka} Here the
$x=0$ compound, ${\rm Cu Br}_{2}(\gamma-{\rm pic})_2$, is a
homogeneous $S=1/2$ antiferromagnetic Heisenberg chain in which the
${\rm Cu-Cu}$ bond is bi-bridged by two ${\rm Br}$ atoms: ${\rm Cu} <
{ {\rm Br} \atop {\rm Br}} > {\rm Cu}$ and the coupling constant is
given by $J''=20.3{\rm K}$.\cite{ajiro} The $x=1$ compound, ${\rm Cu
Cl}_{2}(\gamma-{\rm pic})_2$, is a bond alternating $S=1/2$
antiferromagnetic Heisenberg chain in which two kinds of ${\rm Cu} < {
{\rm Cl} \atop {\rm Cl}} > {\rm Cu}$ bonds alternate along the
chain.\cite{GJW82} The bond alternation is induced by a freezing
transition of the rotational motion of the methyl-group at
$50K$.\cite{GJW82} The experimentally measured coupling strengths are
$J=13.2{\rm K}$ and $J \alpha$, with $\alpha=0.6$. In the mixed
compound with a small finite concentration of Br ($x \ne 1$), atoms
connected with alternating ${\rm Cu} < { {\rm Cl} \atop {\rm Cl}} >
{\rm Cu}$ bonds form a cluster and different clusters are separated by
bonds with ${\rm Br-Br}$ and/or ${\rm Br-Cl}$ bridges.  The strength
of the ${\rm Cu} < { {\rm Br} \atop {\rm Cl}} > {\rm Cu}$ bond has
been estimated from the theoretically calculated magnetization curve
as $J'=1.3J$.\cite{hida03} In ref.~\citen{ajiro} the data for the
low-temperature susceptibility of the diluted system 
show an algebraic temperature dependence, $\chi
\sim T^{\beta-1}$, which is compatible with the expected behavior in 
the Griffiths phase as given by eq.(\ref{chi_rd}). For a wide range of
the concentration, $x$, the measured effective exponent is about
$\beta_{\rm exp} \approx 0.5 - 0.67$, and shows only a weak
concentration dependence.

\section{Theoretical Model}
A theoretical model for ${\rm Cu Cl}_{2x}{\rm Br}_{2(1-x)}(\gamma-{\rm
pic})_2$ was presented and numerically studied in
ref.~\citen{hida03}. It is not obvious from the information one can
extract from the experiments whether the rotational order of the
methyl-group remains long-ranged in the presence of the dilution by
${\rm Cu}< { {\rm Br} \atop {\rm Cl}} >{\rm Cu}$ or ${\rm Cu} < { {\rm
Br} \atop {\rm Br}} > {\rm Cu}$. Therefore two models were introduced in
\cite{hida03}: i) The {\it fixed parity} model in which the rotational
order is assumed to be perfectly long ranged, i.e.  the $J$ ($\alpha
J$) bonds stay in the same parity position, say at $i={\rm even}$
($i={\rm odd}$), in any bond-alternating cluster.  Introducing a
parity parameter, $p_i$, given by $p_i=1(-1)$ if the ${\rm
Cu} < { {\rm Cl} \atop {\rm Cl}} > {\rm Cu}$ bond has a value of $J$
($\alpha J$), one can describe the parity correlations in the fixed
parity model as $p_i p_{i+2l}=1$ and $p_i p_{i+2l-1}=-1$; ii) The
{\it random parity} model in which the rotational long-range order between
two clusters of alternating bonds is assumed to be completely
destroyed by the dilution. In this case, if $i$ and $i+2l$ ($i+2l-1$)
refer to different clusters, the average parity correlations 
vanish: $[p_i p_{i+2l}]_{\rm av}=[p_i p_{i+2l-1}]_{\rm av}=0$.

In this paper we consider a more general model in which rotational
long-range order is partially destroyed by dilution, so that
correlations between parities in two different clusters are given
asymptotically as $[p_i p_{i+2l}]_{\rm av} = - [p_i p_{i+2l-1}]_{\rm
av}\sim l^{-\rho}$. Here we recover the fixed parity and the random
parity models in the limits $\rho=0$ and $\rho \to \infty$,
respectively. We note that some aspects of the effect of correlated
disorder on quantum systems is studied in ref.~\citen{ri99}. Here we
summarize the values of the coupling constant in eq.(\ref{hamilton})
in the following way:
\begin{equation}
J_i=\left\{ \begin{array}{lll}
J \alpha^{(1-p_i)/2} & \mbox{with prob. $x^2$}\\
J' & \mbox{with prob. $2x(x-1)$}\\
J'' & \mbox{with prob. $(x-1)^2$}
\end{array}
\right.\;.
\label{couplings}
\end{equation}
The parity correlations within a cluster of alternating bonds are
given by $p_i p_{i+j}=(-1)^j$ and the parity correlations for bonds in different clusters 
are defined above for the different models.

The theoretical results for the low-energy behavior of the fixed 
parity and the random parity models obtained from the density 
matrix renormalization group (DMRG) method\cite{hida03} are not fully 
consistent with the measured low-temperature properties of the 
${\rm Cu Cl}_{2x}{\rm Br}_{2(1-x)}(\gamma-{\rm pic})_2$.\cite{ajiro} 
In ref.~\citen{hida03} the random parity model is found 
to be in the random singlet phase, in which the susceptibility exponent is 
$\beta=0$ and is much lower than the experimental value 
$\beta_{\rm exp} \approx 0.5 - 0.67$. For the fixed parity model the 
susceptibility exponent is found to be too large: $\beta \ge 1$.
Another problem with the latter model is that the DMRG analysis 
could not be performed for $x \le 0.4$  due to strong finite size effects.   

In the present paper we revisit the models of the compound ${\rm Cu
Cl}_{2x}{\rm Br}_{2(1-x)}(\gamma-{\rm pic})_2$.  Our study is
different from ref.~\citen{hida03} in two respects. Firstly, we
consider a more general model in which the effect of disorder
correlations are taken into account. Secondly, we use a numerical
implementation of the SDRG method.  This method usually gives very
accurate results for the form of singularities, in particular for the
dynamical exponent\cite{lkir}. This method has been successfully
applied to clarify the low-energy singularities of more complicated
systems, such as random spin ladders\cite{ladder} and two- and
three-dimensional random Heisenberg antiferromagnets\cite{2d}. The major
advantage of this method is that one can consider large systems $L
\sim 1000-2000$ with good statistics compared with the DMRG method.

\section{Numerical results}
\subsection{Numerical method}
The SDRG method for random AF spin chains proceeds as follows: During
the renormalization process, one first identify the strongest bond,
say $J_{23}=\Omega$, which connects sites $2$ and $3$.  If the
neighboring bonds are much weaker, $J_{23} \gg J_{12}, J_{34}$, then
the spins $2$ and $3$ form an effective singlet and can be decimated
out. The new coupling between the remaining sites, $1$ and $4$, is
obtained within a second-order perturbation calculation as:
\begin{equation}
\tilde{J}=\frac{1}{2} \frac{J_{12} J_{34}}{J_{23}}\;.
\label{Jtilde}
\end{equation}
By repeating this decimation process, we gradually reduce the energy
scale, $\Omega$.  At the fixed point, the energy scale is give by
$\Omega \sim L^{-z}$ if only a small fraction $1/L$ of sites is active
(i.e. not yet decimated).  In our computations we consider
finite periodic chains with $L={\rm even}$ sites and decimate until
the last pair of sites having a gap $\Delta \sim \Omega$. From the
distribution of the logarithms of the gap:
\begin{equation}
P(\ln \Delta) \sim \Delta^{1/z}, \quad \Delta \to 0\;,
\label{z}
\end{equation}
we obtain the the dynamical exponent, $z$.  In the random singlet
phase, the dynamical exponent is formally infinite, as described in
eq.(\ref{gap_rs}), and the appropriate scaling combination in a finite
system is given by:
\begin{equation}
\ln\left[ L^{\psi} P(\ln \Delta)\right] \simeq f(L^{-\psi}\ln \Delta)\;.
\label{psi}
\end{equation}
The SDRG method as outlined above is expected to provide asymptotically exact results both in the RS phase\cite{fisher} and
in the Griffiths region\cite{ijl01}.

\begin{figure}[htbp]
 {\par\centering \resizebox*{0.4\textwidth}{!}
 {\includegraphics{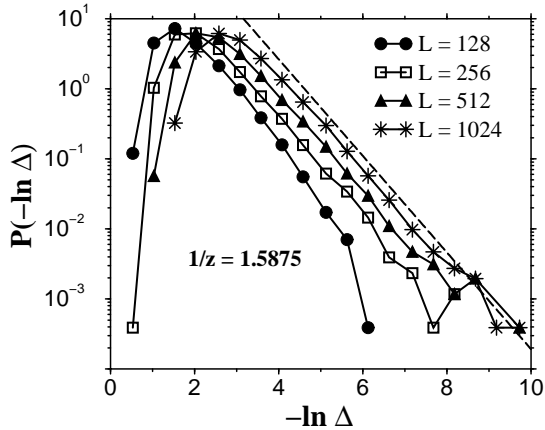}} \par} \vspace{0.5cm} \caption{
 \label{fig:xx} Test of the SDRG method on random dimerized
 $XX$-chains, with $\alpha=0.6, x=0.6$. The slope of the distribution
 of the gaps in a log-log scale agrees well with the known exact
 result for the dynamical exponent, which is given by the slope of the
 broken straight line.}
\end{figure}

In the present model the randomness is discrete and at the starting
point of the renormalization there are several couplings with the same
value of $\Omega$. In this degenerate situation we randomly choose the
actual coupling to be decimated. After a sufficiently large number of
renormalization steps we will have a (quasi)continuous
distribution. To illustrate the correctness of the SDRG procedure for
discrete randomness, we consider the random dimerized
$XX$-chain\cite{XX} in which the couplings take the value $J$ or
$\alpha J$ and a fraction, $x$, of the odd (even) couplings are $J$
($\alpha J$).  To be close to our model in eq.(\ref{couplings}) we
took $\alpha=0.6$ and $x=0.6$ for the $XX$-chain.  In
Fig. \ref{fig:xx} the distribution of the gaps for different finite
systems is shown in a log-log scale, the asymptotic slope of which
agrees very well with the known {\it exact} result\cite{ijr00} for
$1/z$ which is given by the formula $[(J_{\rm e}/J_{\rm o})^{(1/z)}]=1$,
i.e. by the solution of the equation  $(xJ^{(1/z)}+(1-x)(\alpha J)^{(1/z)}) \cdot ((1-x)J^{(-1/z)}+x(\alpha
J)^{(-1/z)})=1$.

\begin{figure}[htbp]
 {\par\centering \resizebox*{0.48\textwidth}{!}
 {\includegraphics{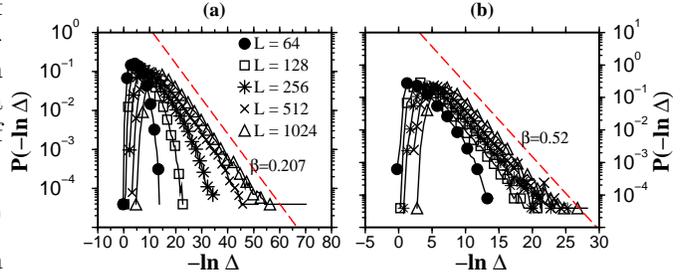}} \par} \vspace{0.5cm} \caption{
 \label{fig:fixed} Distribution of the energy gap for the fixed parity
 model for two concentrations:(a) $x=0.6$, (b) $x=0.8$. The asymptotic
 slope of the curves, indicated by broken straight lines, gives the
 susceptibility exponent, $\beta$.  }
\end{figure}

\begin{figure}[htbp]
 {\par\centering \resizebox*{0.4\textwidth}{!}
 {\includegraphics{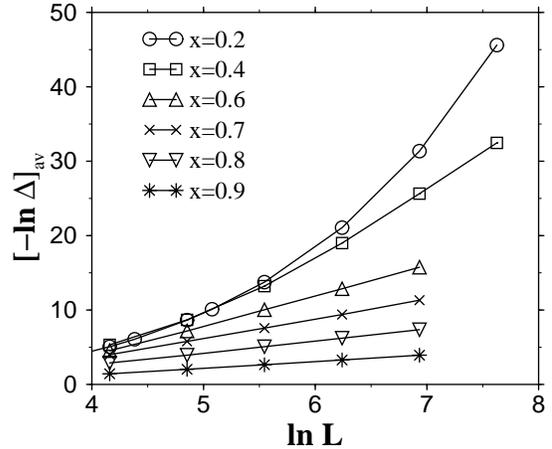}} \par} \vspace{0.5cm} \caption{
 \label{fig:fix_lnE} Average gap of the fixed parity model with
 different concentrations, $x$, as a function of the size, $L$. The
 asymptotic slope of the curves in log-log scale defines the dynamical
 exponent, $z=1/\beta$. Note the strong finite size corrections: for
 sizes used in DMRG ($\ln L \le 5$) we are not in the asymptotic
 regime.  }
\end{figure}

\subsection{Fixed parity model}

\begin{figure}[htbp]
 {\par\centering \resizebox*{0.4\textwidth}{!}
    {\includegraphics{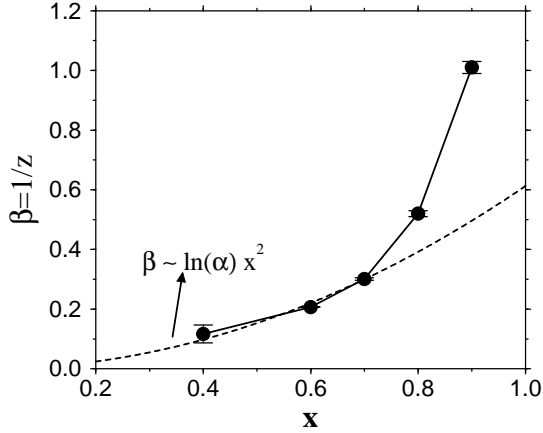}} \par}
 \vspace{0.5cm}
 \caption{
 \label{fig:z_x}
The susceptibility exponent, $\beta=1/z$, for the fixed parity model
as a function of the concentration. For small $x$ the expected
quadratic dependence is denoted by the dotted curve.
 }
\end{figure}

Next, we turn to our model defined in eq.~(\ref{couplings}) and start
with the fixed parity case.  The distribution of the gaps for two
different values of the concentration, $x=0.6$ and $0.8$, are shown in
Fig.~\ref{fig:fixed}. Evidently, the exponent $\beta$, given by the
asymptotic slope of the distributions, is finite and concentration
dependent. It can also be extracted from the scaling behavior of the 
average gap via $[\ln \Delta]_{\rm av} \simeq {\rm const} +
\beta^{-1} \ln L$. As seen in Fig.~\ref{fig:fix_lnE} one can obtain a
reliable estimate of $\beta$ for $x \ge 0.4$. For smaller
concentrations, even the largest system size $L=2048$ is not yet in
the asymptotic regime. The estimates for $\beta$ are depicted in
Fig.~\ref{fig:z_x}. They are systematically lower than the DMRG
results in ref.~\citen{hida03}´, which is probably due to finite-size
effects. Using the DMRG method, ref.~\citen{hida03} reports results up
to $\ln L\le 5$, which is, even for comparatively large
concentrations, not in the asymptotic regime according to
Fig.~\ref{fig:fix_lnE}.
%even for comparatively large concentrations.

As can be seen in Fig.~\ref{fig:z_x}, for small concentrations there
is a quadratic dependence: $\beta \sim x^2$, which can be understood
as follows: In the fixed parity model the average value of the
log-couplings in the odd and even positions are different, so that the
dimer control parameter in eq.~(\ref{delta}) is $\delta \sim x^2 \ln
\alpha \ne 0$.  For small $x$, thus for small $\delta$ we can use
eq.~(\ref{z_delta}), which is compatible with the observed quadratic
$x$-dependence of $\beta$.

\subsection{Random parity model}

\begin{figure}[htbp]
 {\par\centering \resizebox*{0.48\textwidth}{!}
 {\includegraphics{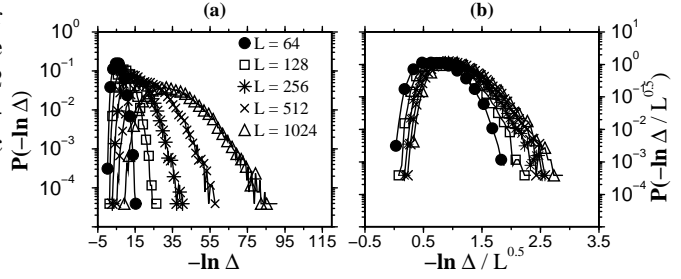}} \par} \vspace{0.5cm} \caption{
 \label{fig:random} Probability distribution of the gaps in the random
 parity model with $x=0.8$ (a). Scaling collapse for different sizes
 using the scaling form in eq.(\ref{psi}) (b).  }
\end{figure}

For the random parity model the gap distribution for an intermediate 
concentration, $x=0.8$, is shown in Fig.~\ref{fig:random}(a).  One
observes that the distribution gets systematically broader with increasing
system size, which is a characteristic of the RS phase.  Indeed, the
logarithmic scaling in eq.~(\ref{psi}) is well satisfied with
$\psi=1/2$, as illustrated in Fig.~\ref{fig:random}(b).  The same
logarithmic scaling with $\psi=1/2$ holds for other values of the
concentration, too.  These results can be understood by noting that
odd and even couplings have on average of the same strength in the random
parity model, consequently the (dimer) control parameter is
$\delta=0$. The short-range correlations in the dimerized sequences,
especially for large concentration, does not modify this large-scale
asymptotic behavior. Our results for the random parity model are in
agreement with the DMRG calculations in ref.~\citen{hida03}.

\begin{figure}[htbp]
 {\par\centering \resizebox*{0.5\textwidth}{!}
 {\includegraphics{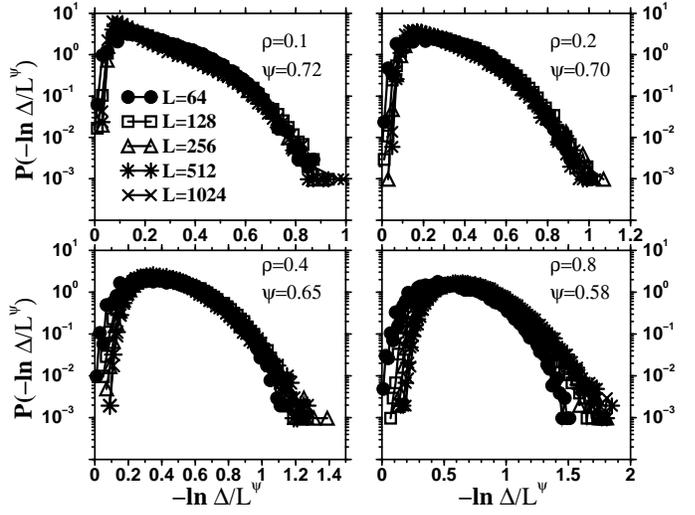}} \par} \vspace{0.5cm} \caption{
 \label{fig:correlated} Scaling plot of the gap distribution for the
 correlated random parity model for different values of $\rho < 1$.
 The $\psi$ exponent is $\rho$-dependent and well described by the
 relation in eq.(\ref{psi_rho}).  }
\end{figure}

\begin{figure}[htbp]
 {\par\centering \resizebox*{0.4\textwidth}{!}
 {\includegraphics{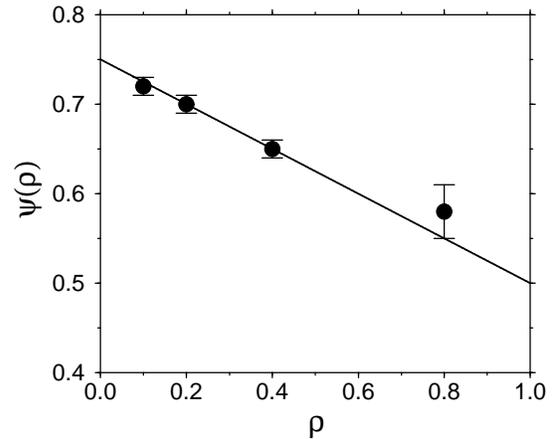}} \par} \vspace{0.5cm} \caption{
 \label{fig:psi_rho} Variation of the estimated exponents $\psi(\rho)$ 
   for the correlated random parity model with the correlation
   parameter $\rho$. The solid line represents the linear fit in Eq.(\ref{psi_rho}).
 }  
\end{figure}

\subsection{Correlated random parity model}

Now we discuss the results for the correlated random parity model. We
recall that according to scaling arguments\cite{WH,ri99} the disorder
correlations modify the critical behavior in the RS phase, provided
the decay exponent is sufficiently small: $\rho < 2/\nu$, where
$\nu=2$ is the correlation length exponent for uncorrelated disorder.
To illustrate this relation we estimate the control parameter,
$\delta$, in a finite chain of length, $L$.  Introducing the notation
$\epsilon_i=\ln(J_{2i}/J_{2i-1})$, we have $\delta^2=1/4
L^2\left[(\sum_{i=1}^{L-1} \epsilon_i)^2\right] \sim 1/L \int_1^L G(r)
{\rm d} r \sim L^{-\rho}$,
where $G(r)$ is the disorder correlator.  Thus $\delta \sim
L^{-\rho/2}$, for $\rho < 1$, whereas for $\rho \ge 1$ we have the
non-correlated disorder result, $\delta \sim L^{-1/2}$.  We can
conclude from this consideration that for $\rho <1$ the correlated
random parity model has a vanishing gap and is in the RS phase,
however with $\rho$-dependent properties. The scaled gap distribution
for four different correlation parameters, $\rho < 1$, is shown in
Fig.~\ref{fig:correlated}, in which the logarithmic scaling collapse
in eq.~(\ref{psi}) can be obtained with different exponents,
$\psi(\rho)$. A plot of the $\psi(\rho)$ data is given in 
Fig.~\ref{fig:psi_rho} in which the $\rho$ dependence can be well fitted with 
the approximate formula:
\begin{equation}
\psi(\rho)=\frac{3 - \rho}{4}, \quad \rho < 1\;.
\label{psi_rho}
\end{equation}
This result is exact at $\rho=1$, in which case the standard, non-correlated RS phase result, $\psi=1/2$ is recovered.
For a small $\rho$ the exponent approaches $\psi \approx 3/4$, but at $\rho=0$ we
jump to the fixed parity model. The linear dependence of $\psi(\rho)$ on
$\rho$ in eq.(\ref{psi_rho}) is in the same form as in the exact
expression for the $XX$-model\cite{ri99} with correlated randomness.

The low-temperature singularity of the susceptibility of the
correlated random parity model is given by the formula:
\begin{equation}
\chi(T) \sim \frac{1}{T (\ln T)^{1/\psi}}\;,
\label{chi_corr}
\end{equation}
in which $\psi$ is $\rho$-dependent, as given in Eq.(\ref{psi_rho}).

\section{Discussion}
We close our paper by comparing the experimentally measured
low-temperature susceptibility of the compound ${\rm Cu Cl}_{2x}{\rm
Br}_{2(1-x)}(\gamma-{\rm pic})_2$ with the results of our theoretical
calculations. We recall that the measured susceptibility exponent is
finite, $\beta_{\rm exp} \approx 0.5-0.67$, and has only a weak
concentration dependence. These properties are partially compatible
with the results for the fixed parity model, which is found to be in
the Griffiths phase, where $\beta$ is finite for all values of the
concentration. According to Fig.~\ref{fig:z_x} there is a range of
concentration around $x \approx 0.8$ in which our results for $\beta$
agree with the experimental values. However, this range is rather
narrow and our values for $\beta$, which have a substantial
concentration dependence, are generally significantly smaller than
$\beta_{\rm exp}$. For models with non-strictly correlated parities,
in which case $\beta$ is found formally zero, the agreement with the
experiment is even less satisfactory. One possible explanation of this
discrepancy between experiment and theory is that the experimentally
measured $\beta_{\rm exp}$ are effective, temperature dependent
values, which should approach the true behavior as $T \to 0$. Indeed,
as seen in Fig.~\ref{fig:fix_lnE} the local slopes of the curves that
show $\ln \Delta$ vs. $\ln L$, which define the effective exponent
$z(L)=1/\beta(L)$, have a strong size dependence. This can be
converted into a temperature dependence through $\Omega \sim T \sim
L^{-z}$. The corrections to $\beta(L)$ are particularly strong for
small values of $x$. For moderately large sizes, $\ln L \approx 5$,
the effective exponents have only a weak concentration dependence. The
leading finite size corrections are of the form, $\beta(L) \simeq
\beta + a/L$, which are compatible with a temperature correction as:
\begin{equation}
\beta(T) \simeq \beta + c T^{\beta} + \dots.\quad \beta>0. \;,
\label{beta_fixed}
\end{equation}
In the random singlet phase with $\beta=0$, both for correlated and
non-correlated parity, the effective exponents have a logarithmic
temperature dependence:
\begin{equation}
\beta(T) \simeq  \frac{c_1}{|\ln T|} + \dots.\quad \beta=0. \;,
\label{beta_rs}
\end{equation}
which can be obtained by analyzing eqs.(\ref{chi_rs}) and
(\ref{chi_corr}). Indeed, these finite temperature corrections are
strong for the fixed parity model, particularly for small $\beta$,
i.e. for a small concentration, $x$.

\begin{figure}[htbp]
 {\par\centering \resizebox*{0.45\textwidth}{!}
 {\includegraphics{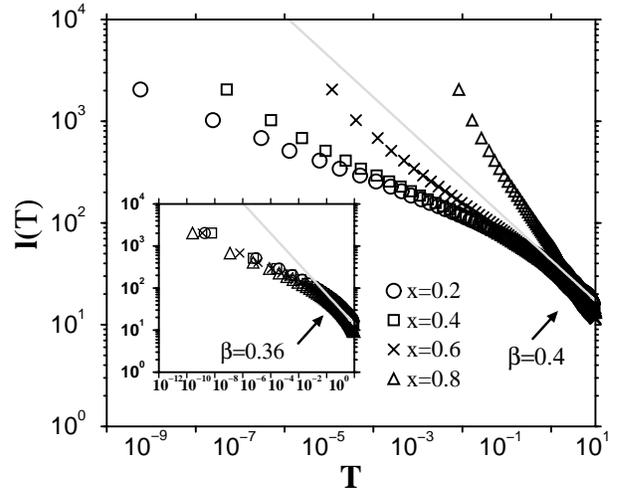}} \par} \vspace{0.5cm} \caption{
 \label{fig:beta_T} Temperature dependent length-scale, $l(T)$, which
 is obtained by performing the renormalization down to an energy
 $\Omega=T$, for the fixed parity and the random parity (inset)
 models. In the log-log plot the local slope of the curves gives the
 temperature dependent effective exponent, $\beta(T)$. The slope of
 the straight line indicates the effective exponent at $T=2K$.  }
\end{figure}

We tried to estimate the effective exponent, $\beta(T)$, at the
temperature of the experimental measurement $T=2 {\rm K}$. For this we
performed the renormalization transformation down to an energy-scale,
$\Omega=T$, and measured the fraction of non-decimated sites:
$n(\Omega)=1/l(\Omega)$. Here the length-scale is given by $l(\Omega)
\sim \Omega^{-\beta}$. Our results for $l$ as a function of the temperature
%$\Omega=TMO45596$
for the fixed parity and the random parity model are presented in
Fig. \ref{fig:beta_T} for different values of the concentration. In
the log-log plot the local slope of the curves is just the effective
exponent, $\beta(T)$, at $T=\Omega$. As can be seen in this figure the 
effective exponent, $\beta(T)$, approaches its limiting value only if the 
temperature is sufficiently low. At the temperature of the measurement, 
$T=2{\rm K}$, the asymptotic region seems to be still quite far. 
For the fixed parity model the effective exponent is about $\beta=0.4$, 
which is practically independent of the concentration. This result is
consistent with the experimental results of ref.~\citen{ajiro}. For
the random parity model, as shown in the inset, the effective exponent
continuously vary with the temperature. Its value at $T=2 {\rm K}$ is
about $\beta=0.36$, which is also consistent with the
experiments. Therefore, using the available experimental data it is
not possible to distinguish the type of parity correlations present in
the compound, ${\rm Cu Cl}_{2x}{\rm Br}_{2(1-x)}(\gamma-{\rm pic})_2$
and the question, which type of model should be used for its
theoretical description remains still open. Experimental measurements
at lower temperatures are needed to clarify this point.

\section*{Acknowledgment}
This work has been supported by a German-Hungarian
exchange program (DAAD-M\"OB), by the Hungarian National Research Fund
under grant No OTKA TO34138, TO37323, MO45596 and M36803, by the
Ministry of Education under grant No. FKFP 87/2001 and by the Centre
of Excellence ICA1-CT-2000-70029.  
%Numerical calculations are
%partially performed on the Cray-T3E at Forschungszentrum J\"ulich.
%\appendix
%\section{Sample}

%Equations in the appendix will be numbered as (A$\cdot$1), (A$\cdot$2), (A$\cdot$3) \ldots.


\begin{thebibliography}{99}
\bibitem{XX}
        For the random $XX$ model the decimation rule in Eq.(\ref{Jtilde}) is modified
        by having a prefactor $1$ instead of $1/2$.
\end{thebibliography}
\end{document}